# Spin-orbit coupling effects on spin-phonon coupling in $Cd_2Os_2O_7$

Taehun Kim[1,2,3], Choong H. Kim[2,3], Jaehong Jeong[2,3], Pyeongjae Park[1,2,3], Kisoo Park[2,3], Ki Hoon Lee[2,3,4,5], Jonathan C. Leiner[2,3,6], Daisuke Ishikawa[7], Alfred Q. R. Baron[7], Zenji Hiroi[8], and Je-Geun Park[1,2,3,#]

[1]Center for Quantum Materials, Seoul National University, Seoul 08826, Republic of Korea

[2]Center for Correlated Electron Systems, Institute for Basic Science, Seoul 08826, Republic of Korea

[3]Department of Physics and Astronomy, Seoul National University, Seoul 08826, Republic of Korea

[4]Center for Theoretical Physics of Complex Systems, Institute for Basic Science, Daejeon 34126, Republic of Korea

[5]Department of Physics, Incheon National University, Incheon 22012, Republic of Korea

[6]Physik-Department, Technische Universität München, D-85748 Garching, Germany

[7]Materials Dynamics Laboratory, RIKEN SPring-8 Center, Hyogo 679-5148, Japan

[8]Institute for Solid State Physics, The University of Tokyo, Chiba 277-8581, Japan

#corresponding author: jgpark10@snu.ac.kr

**Abstract**

Spin-orbit coupling (SOC) is essential in understanding the properties of $5d$ transition metal compounds, whose SOC value is large and almost comparable to other key parameters. Over the past few years, there have been numerous studies on the SOC-driven effects of the electronic bands, magnetism, and spin-orbit entanglement for those materials with a large SOC. However, it is less studied and remains an unsolved problem in how the SOC affects the lattice dynamics. We, therefore, measured the phonon spectra of $5d$ pyrochlore $Cd_2Os_2O_7$ over the full Brillouin zone to address the question by using inelastic x-ray scattering (IXS). Our main finding is a visible mode-dependence in the phonon spectra, measured across the metal-insulator transition at 227 K. We examined the SOC strength dependence of the lattice dynamics and its spin-phonon (SP) coupling, with first-principle calculations. Our experimental data taken at 100 K are in good agreement with the theoretical results obtained with the optimized U = 2.0 eV with SOC. By scaling the SOC strength and the U value in the DFT calculations, we demonstrate that SOC is more relevant than U to explaining the observed mode-dependent phonon energy shifts with temperature. Furthermore, the temperature dependence of the phonon energy can be effectively described by scaling SOC. Our work provides clear evidence of SOC producing a non-negligible and essential effect on the lattice dynamics of $Cd_2Os_2O_7$ and its SP coupling.

## 1. Introduction

Spin-orbit coupling (SOC) is a central concept to understanding the diverse phenomena and several strong correlation effects of 4$d$ and 5$d$ transition metal oxides [1]. Depending on the strength of SOC, both the electronic and magnetic structures of a specific material are modified, and several distinctive quantum phases can emerge as a result. For example, when the spin and orbital degrees of freedom are strongly coupled, a spin-orbital entangled state becomes the ground state, as found in iridates [2]. Since 5$d$ transition metals have considerable SOC energies, materials having Ir or Os atoms have been the focus of the recent extensive studies.

$Cd_2Os_2O_7$ is one of the candidate materials having strong SOC, and it exhibits several phenomena exciting in their own right. The crystal structure of $Cd_2Os_2O_7$ has a pyrochlore lattice with the space group of Fd3-m, and the lattice parameter is $a$ = 10.1604 Å [3]. Os atoms mainly govern the magnetism with a 5$d^3$ electron configuration with $S$ = 3/2. Unlike iridates, the electronic ground state is not spin-orbital entangled and therefore the effect from SOC is less straightforward, as shown by the absence of spin-orbit exciton found in the resonant inelastic x-ray scattering (RIXS) study [4]. It is also known that the magnetic structure has the so-called all-in-all-out (AIAO) state [4,5] with the Néel temperature of $T_N$ = 227 K, as shown in Fig. 1(a). The magnetic transition is also accompanied by a drastic change in the resistivity and a metal-insulator transition (MIT) at almost the same temperature [6]. While the mechanism of the MIT in $Cd_2Os_2O_7$ was first suggested as a Slater-type [7], it was later re-considered that the MIT is a rare realization of the Lifshitz-type [8,9].

The first-principle [10] and many-body quantum-chemistry calculations [11] have so far been carried out to understand how the AIAO ordering state becomes stable, especially given its sizeable magneto-crystalline anisotropy. In these theoretical studies, a sizeable single-ion anisotropy (SIA) is found along the local [111] axis, and there is even the trigonal distortion of the $OsO_6$ octahedron. The large SIA is experimentally confirmed, too [4,12,13]. A spin Hamiltonian with a large anisotropy was used to analyze the multiple spin-flip excitations observed in RIXS experiments [4,12] and the two-magnon continuum from Raman spectroscopy [13]. The spin Hamiltonian used for the proceeding analysis can be written as

$$H_{\text{spin}} = J \sum_{<ij>} S_i \cdot S_j + \sum_{<ij>} D_{ij} \cdot S_i \times S_j + K \sum_i (S_i \cdot n_i)^2$$

, where $J$, $D$, and $K$ are the Heisenberg interaction, the Dzalyoshinskii-Moriya interaction, and the SIA, respectively. $n_i$ is the unit vector pointing to the center of Os tetrahedron, i.e., the local [111] axis. The best fit to the two-magnon signals observed in Raman spectra was obtained with $J$ = 5.1, $K$ = -5.3 meV, and the $D$ value of about 1/3 of $J$ [13].

Such a large SIA is essential to understanding the phonons of $Cd_2Os_2O_7$ as well. In $Cd_2OS_2O_7$, some IR-active phonon modes display a large renormalization upon entering the magnetic transition, which was later taken as evidence of the so-called spin-phonon (SP) coupling [14]. While a usual SP coupling appears in the form of the exchange-striction [15], magneto-striction [16,17], or magnon-phonon hybridization [18–21], the SP coupling of $Cd_2Os_2O_7$ is somewhat unique as the main driving force is the SIA. Interestingly, SIA was considered as a possible coupling between phonon and magnon, more precisely crystal field excitations, for rare-earth-based magnetic materials [22]. Note that the more common nomenclature for such effect is a spin-lattice coupling with notable examples [23], which were even considered in the seminal work in the 50s by Kittel [24]. However, we have chosen to stick with the SP coupling in this work as it is in more common use concerning $Cd_2Os_2O_7$. According to the current understanding, the extensive modification of the crystal field induces the SP coupling when the atoms are vibrating [14]. It implies that SOC has an essential role in the SP coupling of $Cd_2Os_2O_7$ since the SIA is originated from the crystal field generated by the surrounding environment of an $OsO_6$ octahedron. Therefore, it is an interesting problem to study the effect on the phonon modes due to the SOC strength theoretically and compare these theoretical results with the experimental data.

Despite that there have been numerous studies on the SOC-related electron structure or magnetism in Ir or Os based materials, studies on the relation between SOC and the lattice dynamics are relatively rare. For this lack of the otherwise important information, we decided to investigate the full spectrum of phonons of $Cd_2Os_2O_7$ over a wide moment-energy range. We used both the inelastic x-ray scattering (IXS) technique and the first-principle calculations. Experimentally, we succeeded in mapping the phonon dispersion over the full Brillouin zone and subsequently confirming that the SP coupling exists in a wide range of momentum and energy transfers. In the subsequent theoretical studies using the density-functional theory (DFT) calculations, we demonstrate that SOC is the main ingredient of the SP coupling by examining the phonon spectra while scaling SOC strength and controlling electron correlation U.

## 2. Experimental details and theoretical methods

High quality $Cd_2Os_2O_7$ single crystals were synthesized by the chemical transport method [8]. Characterizations of the crystals using the single-crystal x-ray diffraction and the magnetization measurements were performed, as shown in Fig. 2. The magnetic ordering temperature observed in the susceptibility data is found to be $T_N \sim 227$ K, which is the same as the reported one [8]. The IXS experiments were carried out at the Quantum Nano Dynamics beamline, BL43LXU [25], of the RIKEN SPring-8 Center. Backscattering using the silicon (11 11 11) reflection was used to attain an energy resolution of about 1.5 meV at an x-ray energy of 21.75 keV. A 2-dimensional array of analyzers was used for the parallel collection of data at 23 different momentum transfers [26]. The analyzer acceptance was set to 40 × 40 mm by slits at 8.9 m from the sample position corresponding to a momentum resolution of about (0.048, 0.011, 0.049) Å$^{-1}$. Based on the observed (6 6 6) Bragg peak, we confirmed that the full-width at half-maximum (FWHM) of the peak is 0.04°, which is the evidence of our samples' high quality.

The first-principles calculations were performed using DFT+U with the PBEsol exchange-correlation functional implemented in VASP [14]. The AIAO magnetic configuration was considered within a noncollinear DFT formalism. The frozen phonon method was used to calculate the phonons with U = 0.5, 1.0, 1.5, 2.0 eV chosen to consider the local Hubbard interaction in Os atoms by employing PHONOPY [27]. We varied the scaling factor of SOC from 0.4 to 1.0 in the Hamiltonian to control the SOC strength. Note that the scaling factor 1.0 means that we fully considered SOC in our DFT calculations while we examined the SOC effects by adding additional term to the Kohn–Sham equations used in the DFT calculation. In this approach, by setting the SOC strength to 0.5 we artificially reduced the SOC while carrying out self-consistent DFT calculations.

## 3. Experimental results and discussions

We collected the IXS spectra mainly for the (6 6 6), (6 6 8), and (8 6 6) Brillouin zones at 282 and 100 K: one was collected below the transition temperature of $T_N \sim 227$ K, and another is above it. Figs. 3(a-d) show the phonon spectra taken at four selected ***Q*** points corresponding to the zone boundaries, labeled as shown in Fig. 1(b). All the spectra were measured at 100 K, well below $T_N$. Our data reveal acoustic phonon modes as well as several optical phonon modes up to 35 meV. The solid red lines represent the calculated IXS cross-sections based on the phonon energies and eigenvectors from the DFT calculation. Note that all the calculated cross-sections were corrected for the Bose factor and the Debye-Waller factor before being compared with the experimental results. And the Lorentzian functions were used for the spectral calculations with the fixed linewidth at 1.6 meV, similar to the experimental resolution. After these corrections, the calculated IXS spectra are found in excellent agreement with the data. To obtain the best fit to the experimental results, we chose the U and SOC values in our DFT calculations as U of 2.0 eV and the scaling factor for SOC of 1.0. These values are reasonable and consistent with other studies [9,10]. We also obtained the phonon dispersion curves, as plotted in Fig. 3(e), with an overall good match with the experimental results.

As for the temperature dependence through the magnetic and metal-insulator transition, we found visible energy shifts for some optical modes in our data. Although a similar shift was previously reported from the optical spectroscopy studies [13,14], these previous data intrinsically probe phonons only near the zone center. On the other hand, we measured phonons over the full Brillouin zone, which is essential to shedding light on the full understanding of the SP coupling. One example of the IXS spectra taken at both 282 and 100 K with $Q$ = W (6.50 6.00 7.00) is shown in Fig. 4(a). The three peaks labeled as A, B, and C do not move, whereas the peaks D and E shift towards lower energy when the temperature increases. Since there is no structural transition over the temperature range of interest, these shifts ought to be the direct effect of the SP coupling. There is a possibility that the phonon change may be induced by the change of lattice parameter through the phase transition. Based on the Grüneisen parameter reported in pyrochlore hafnates [28], the phonon energy shift induced by the change of lattice parameter is expected to be $|(E_{300K} - E_{100K})|/E_{100K} \approx 1.5$ %. However, the energy change of the peak D is found to be 4.79 %, which is much larger than 1.5 %, enabling us to rule out the lattice parameter change as the primary factor. Furthermore, the possibility of electron-phonon coupling has been already discussed in the previous study [14], which argued that it is small enough to be neglected [14].

This surprising result suggests that the SP coupling has a mode dependence at finite $Q$, which was never reported before. The phonon modes corresponding to the peaks D and E mainly consist of the vibration of the magnetic Os atoms, according to the mass of Os atoms and the eigenvector analysis from the DFT calculations. We also plotted the IXS spectra taken at the other momentum position near K at (6.60 5.95 6.60) in Fig. 4(d), which shows no visible signs of phonon energy changes. These experimental observations constitute the first and direct evidence of the SP coupling with the visible momentum dependence.

Another noticeable change is the linewidth of the peaks D and E, as shown in Fig. 4(a, d). The FWHM of the peaks D and E measured at 100 K are found to be 1.68 and 1.79 meV, similar to the energy resolution. However, when the temperature gets increased to 282 K above the transition temperature, the experimental linewidth is obtained as 4.59 and 2.56 meV, much larger than expected of thermal broadening effects alone. We note that a similar magnitude of broadening was reported for zone-center phonons by the previous IR study [14], although our data cover a much wider space of momentum and energy. In principle, this broadening can be related to either the electron-phonon coupling or the SP coupling that appeared through the MIT and magnetic transition in $Cd_2Os_2O_7$. Taken together, we think that the broadening observed in both experiments is more consistent with the scenario of the SP coupling.

It is worth noting that the SP coupling explains the excellent agreement between the IXS data taken at 100 K and the current DFT calculations without invoking any other interaction terms. The excellent agreement observed suggests that the DFT calculation captured all the essential physics of the SP coupling in $Cd_2Os_2O_7$. Since the anharmonic or higher-order interactions (e.g., phonon-phonon, or magnon-phonon [18]) are not included in our calculations, there is no expected energy renormalization by the quasiparticle hybridization or the band repulsion in our theoretical results. For the question of magnon-phonon coupling, we note that the magnon modes are located from 60 to 100 meV according to its spin Hamiltonian [13]. And so they would have little chance of interacting with phonons located at a much lower energy of from 10 to 35 meV [21].

Such mode-dependent SP coupling is further demonstrated in the DFT calculations to have nontrivial U and SOC effects. We calculated the IXS cross-sections for the observed spectra for the better comparison between the experimental and theoretical results, as shown in Fig. 4. One can see that the reduction of SOC changes the peaks D and E visibly (Fig. 4(b)), while the reduction of U changes the peaks B, C, D, and E (Fig. 4(c)) instead. So SOC and U have different effects on different phonons. To be more specific, the energy renormalization $\Delta E/E_0$ of 0.42 % is found for peak A by changing temperature from 282 to 100 K, which is defined as $\Delta E/E_0 = |(E_{282K} - E_{100K})|/E_{100K}$. On the other hand, the $\Delta E/E_0$ induced by changing U from 0.5 to 2.0 eV and scaling SOC strength from 0.4 to 1.0 are found to be 2.02 % and 0.63 %, respectively. For peak D, the $\Delta E/E_0$ obtained by changing temperature, U, and SOC are found to be 4.79 %, 8.08 %, and 4.91 %, respectively. So, the observed energy renormalizations can be nicely explained by SOC. In the case of Fig. 4(d), whose peaks do not move in the IXS spectra with changing temperature, one can see that scaling SOC moves only the peak δ (Fig. 4(e)) while changing U moves the peaks β, γ, and δ (Fig. 4(f)). Thus, there are changes to all-optical phonons by varying the U value. On the other hand, we have found that some specific phonons are more sensitive to SOC strength.

We also confirmed that the temperature and SOC dependence are similar in the IR active phonons, as shown in Fig. 5(a). When the temperature or SOC strength decreases, the IR active phonon energies are reduced. We tried to fit the energy of three IR active phonon modes as a function of the temperature by comparing them with the calculated phonon energy shift as a function of SOC strength. Note that the data and the labels for three phonons are adapted from the reference [14]. The red diamonds represent the fitted results that are matched with the three IR phonon energies simultaneously, and they are in good agreement with our data discussed above. We also found an approximately linear relation between SOC and temperature based on the best fit, as shown in Fig. 5(b). This linear relation between SOC and temperature indicates that SOC can effectively capture the temperature dependence of the SP coupling, as seen in the shift of the phonons position.

We caution that although successful in this work, one needs more careful theoretical studies as to our approaches of directly comparing the temperature with SOC because the first-principle calculations are done at zero temperature. And the reduction of SOC would not be realistic because SOC is one of the intrinsic parameters from the material itself, which is independent of the temperature. Instead, we would like to emphasize that what we have shown in this work is that scaling SOC can still be an effective and practical way to interpret the temperature effect in the experimental observations. The phonon energy change $\Delta\omega_\sigma$ caused by the SP coupling is mainly determined by the single-ion anisotropy $K$, and the mean value of the Os spins formulated as below [14].

$$\Delta\omega_\sigma = \sum_i \frac{\partial^2 K}{\partial u_{i,\sigma}^2} \langle(\mathbf{S}_i \cdot \mathbf{n}_i)^2\rangle$$

Here, the $\mathbf{u}_{i,\sigma}$ is an atomic displacement with the atom index $i$ and band index $\sigma$. The second derivative of the $K$ term is related to the SP coupling constant, and it is dependent on the variation of anisotropy when the atoms are vibrated. And the single-ion anisotropy $K$ is proportional to the square of SOC [29]. Scaling SOC strength in the calculations directly perturbs the first term. We think that the reduction of SOC strength decreases the coupling constant, and it gradually reduces the phonon energy, like the experimental observation. In real material, $K$ is a temperature-independent value while $\langle(\mathbf{S}_i \cdot \mathbf{n}_i)^2\rangle$ is a temperature-dependent quantity. As temperature increases above $T_N$, spin fluctuations get decreased and $\Delta\omega_\sigma$ should decrease as well. Therefore, changing $K$ by scaling strength of SOC and changing magnetic moment fluctuation by the temperature have the same effect on $\Delta\omega_\sigma$.

## 4. Conclusion

In summary, we conducted the experimental investigation of phonons over the full Brillouin zone in $Cd_2Os_2O_7$ using IXS to find a clear mode dependence in its temperature effects. The IXS spectra of the SP-coupled phase ($T<T_N$) are well explained by the DFT calculations based on the best parameters of U = 2.0 eV with SOC. By controlling both U and SOC in the DFT calculations, we found that the observed mode-dependent SP coupling is mainly driven by SOC, but not by U, as confirmed by comparison of phonon dispersion from the IXS experiment and the DFT calculation. Furthermore, the SOC can effectively explain the temperature effect of the SP coupling in $Cd_2Os_2O_7$. These results imply that SOC is the key to explaining the SP coupling in $Cd_2Os_2O_7$. Our result suggests that SOC is a fundamental ingredient in understanding phonons in 5$d$ transition metal oxides. The effects of SOC on lattice dynamics shown in this study can be applied further to other compounds with strong spin-lattice coupling, or SP coupling.

## Acknowledgments

Work at the Center for Quantum Materials was supported by the Leading Researcher Program of the National Research Foundation of Korea (Grant No. 2020R1A3B2079375). Work at the IBS CCES was supported by the Research Center Program of Institute for Basic Science (IBS) in South Korea (Grant No. IBS-R009-G1 and IBS-R009-D1). IXS experiments were supported by the RIKEN SPring-8 Center under Proposal No. 20180035.

**Figures**

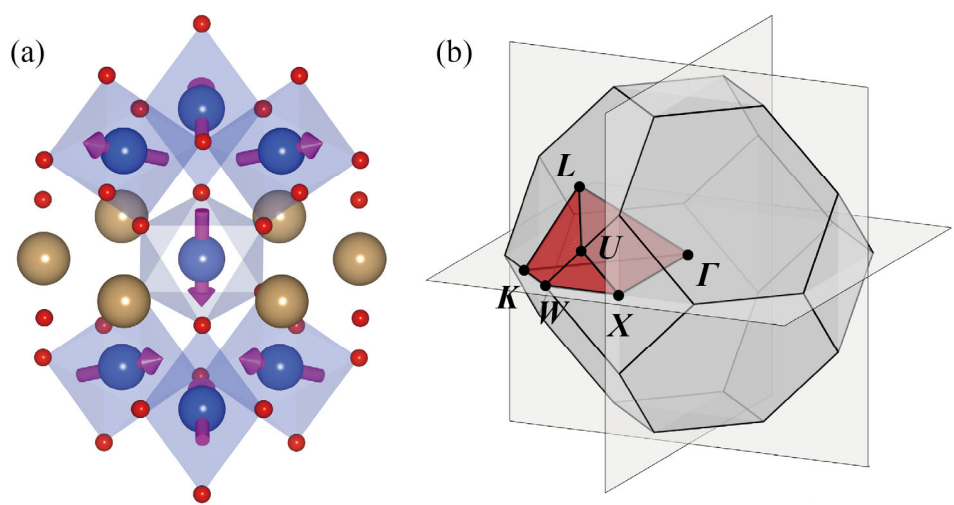

Figure 1. (a) Crystal and magnetic structure of $Cd_2Os_2O_7$. The yellow (blue) spheres indicate the Cd (Os) atoms. The red spheres are the O atoms. Arrows indicate the direction of ordered magnetic moments in Os atoms, which represent the AIAO ordered state. (b) The Brillouin zone of the pyrochlore lattice. Because of the crystalline symmetry of $Cd_2Os_2O_7$, the Brillouin zone can be folded into the red colored irreducible Brillouin zone, which is $1/48^{th}$ of the original Brillouin zone.

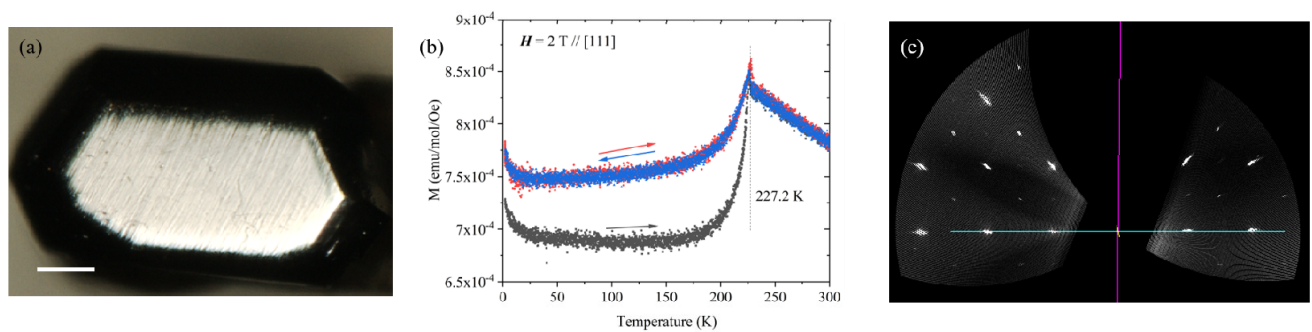

Figure 2. (a) The photo of the single crystal used in the IXS experiment. The scale bar shown in the plot corresponds to 100 μm. (b) The temperature dependence of magnetization is measured under the external field of 2 T. The magnetization measurement was performed using an MPMS3 SQUID magnetometer, Quantum Design. (c) Single crystal X-ray diffraction measurements along the [H0L] plane. The diffraction patterns were collected using XtaLAB P200, Rigaku.

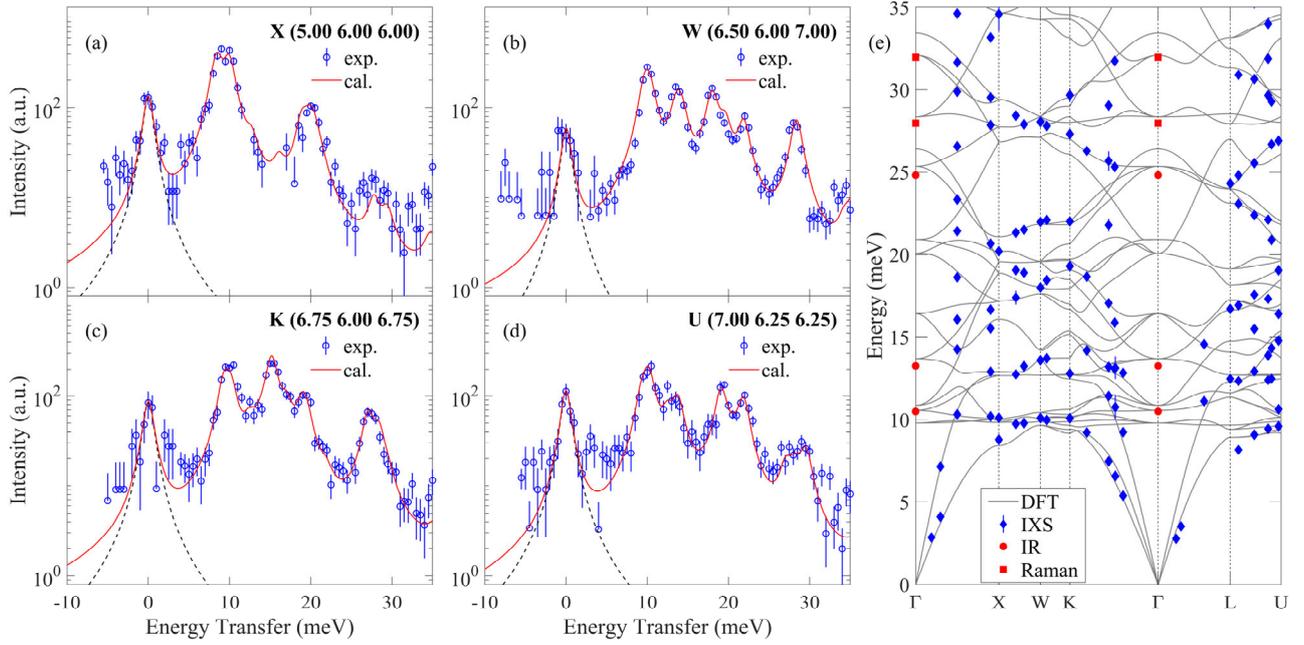

Figure 3. (a-d) The IXS spectra at several ***Q*** points. The blue symbols are the IXS spectra taken at 100 K. The solid red lines indicate the IXS cross-section calculated by the DFT theory with U=2.0 eV with SOC. The Debye-Waller factor and the Bose factor corrections are applied for the results. The black dashed lines are the elastic peak convoluted with the experimental resolution profile. (e) The phonon dispersion curves were obtained from both IXS experiments (blue diamonds) and the calculations (solid gray lines). The IR (red spheres) and Raman (red squares) results are adapted from the references [13,14].

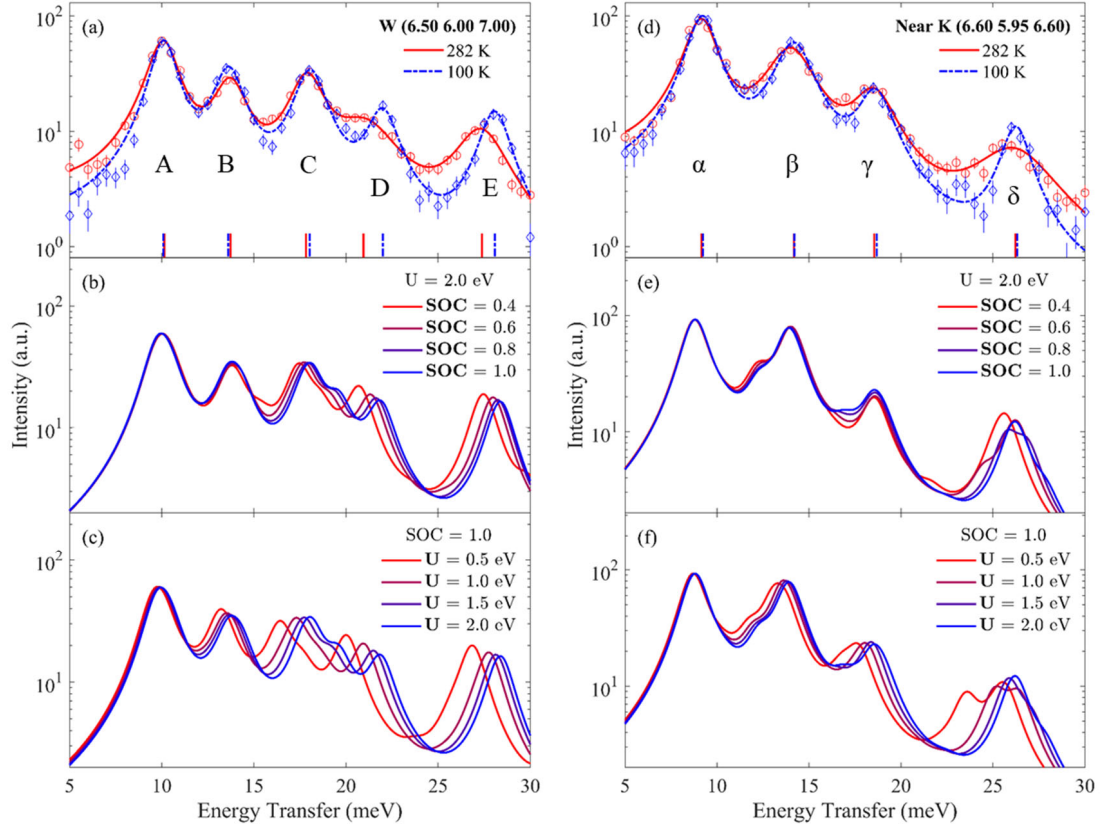

Figure 4. The red and blue symbols are the IXS spectra taken at 282 and 100 K. The **Q** points for each spectrum are (a) W (6.50 6.00 7.00) and (d) near K (6.60 5.95 6.60), respectively. The solid and dashed lines are the fitted lines using multiple Lorentzian functions without any background functions. (b, e) The calculated IXS cross-sections with scaling SOC strength. The numbers shown in the legend represent the scaling factor for SOC strength used in the calculations. (c, f) The calculated IXS cross-sections with different U.

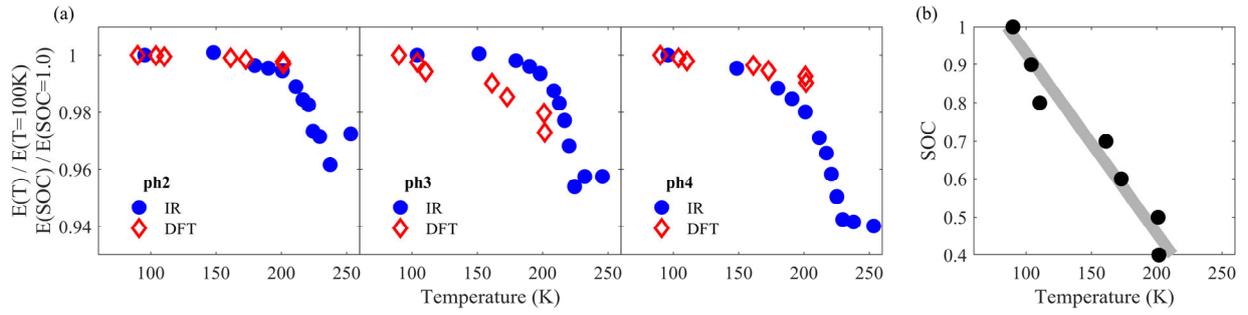

Figure 5. (a) The temperature dependence on the phonon energy of three IR active phonon modes (blue dots) is adapted from reference [14]. The phonon energies for each temperature are normalized to the phonon energy measured at 100 K. SOC dependence of phonon energy (red diamonds) is calculated for three IR active phonon modes. The energies for each SOC strength are normalized to the energy calculated with the scaling factor for SOC of 1.0. We fit the SOC dependence of the phonon energy to match the temperature dependence of three IR active phonon modes simultaneously. (b) The relation between SOC and the temperature obtained from the fitted results, as shown in Fig. 5(a).